\documentclass[11pt,a4paper]{article}
\pdfoutput=1
\usepackage{jheppub}
\usepackage{booktabs}
\usepackage{tabularx}
\usepackage{multirow}
\usepackage{array}
\usepackage{microtype}
\usepackage{float}

\newcommand{\JHEPArxiv}[2]{\href{https://arxiv.org/abs/#1}{arXiv:#1#2}}
\newcommand{\JHEPInspire}[1]{\href{https://inspirehep.net/literature?q=texkeys:#1}{INSPIRE}}

\graphicspath{{figures/}}

\title{\noindent Harmonic-dependent geometry-to-flow transfer in AMPT Ru+Ru and Zr+Zr isobar collisions}
\author[a]{\noindent Murad Badshah}
\author[b]{Muhammad Ajaz}

\affiliation[a]{Department of Physics, Abdul Wali Khan University Mardan, Mardan, Pakistan}
\affiliation[b]{Department of Physics, Faculty of Science, University of Tabuk, Tabuk 47913, Saudi Arabia}

\date{\textit{E-mail:}
Murad Badshah: 
\href{mailto:murad_phy@awkum.edu.pk}{\ttfamily murad\_phy@awkum.edu.pk},
\href{mailto:muradbadshah25295@gmail.com}{\ttfamily muradbadshah25295@gmail.com};\\
Muhammad Ajaz:
\href{mailto:muhammad.ajz@cern.ch}{\ttfamily muhammad.ajaz@cern.ch}}
\abstract{We present a fixed-$N_{\rm part}$ study of geometry-to-flow transfer in string-melting AMPT simulations of $^{96}{\rm Ru}+^{96}{\rm Ru}$ and $^{96}{\rm Zr}+^{96}{\rm Zr}$ collisions at $\sqrt{s_{NN}}=200$ GeV. Four nuclear configurations are considered: deformation-only Ru+Ru, deformation-only Zr+Zr, deformation plus neutron skin Ru+Ru and deformation plus neutron skin Zr+Zr. The AMPT/HIJING initialization is modified to include deformed Woods--Saxon densities and, when enabled, separate proton and neutron Woods--Saxon radii and diffuseness parameters (neutron skin effect). We introduce an eccentricity-normalized isobar response double ratio, $D_n=(v_n^{\rm Ru}/v_n^{\rm Zr})/(\varepsilon_n^{\rm Ru}/\varepsilon_n^{\rm Zr})$, evaluated in common participant-number intervals. This observable eliminates the initial-geometry eccentricity ratio, and investigates whether the final-state flow ratio is fully determined by the initial-geometry ratio. We find that the elliptic double ratio is nearly unity, and the triangular double ratio is consistently greater than unity for the two deformation-only and deformation-plus-skin configurations. The positive $D_3-1$ pattern is maintained when varying the $N_{\rm part}$ binning, peripheral-bin treatment, flow $p_T$ range and rapidity/pseudorapidity acceptance. The result identifies a harmonic-dependent AMPT response: Ru/Zr elliptic flow follows leading eccentricity scaling to high accuracy, whereas triangular flow retains a residual response component after the triangularity ratio is divided out.}

\begin{document}
\maketitle

\section{Introduction}
\label{sec:introduction}

In relativistic heavy-ion collisions, the spatial structure of two colliding nuclei is altered to a final state distribution of produced particles. When two nuclei collide outside of the middle, the overlap resulting in an almond-form is responsible for the phenomena of elliptic flow. Triangular (and higher harmonics) flows arise from event-by-event density fluctuations and higher multipole components of the nuclear shape. The Fourier coefficients $v_n$ of the end azimuthal distribution are therefore among the most convenient observables to link the shape of the initial geometry to the transport properties and collective expansion of the medium produced \cite{Ollitrault:1992bk,Voloshin:1994mz,Poskanzer:1998yz,Heinz:2013th,Gale:2013da}. This relation is typically written in the approximate form $v_n\sim \kappa_n\varepsilon_n$ and is usually understood in terms of an initial eccentricity $\varepsilon_n$ and an effective response coefficient $\kappa_n$ which includes the hydrodynamic response, microscopic transport, hadronic rescattering and analysis-specific choices.

It is not expected that the response relation for all the harmonics is the same. Elliptic flow is strongly linked to the average overlap geometry of the system, while the triangular flow is more fluctuation-driven and more sensitive to the density perturbations, finite size effects and/or nonlinear correlations in the initial-state geometry of the system \cite{Alver:2010gr,Teaney:2010vd,Qiu:2011iv,Schenke:2010rr,Gardim:2011xv}. This harmonic dependence motivates measurements and simulations that do not only ask whether a nucleus is deformed, but also ask how different harmonic components of its geometry survive the dynamical evolution. In this sense the response coefficient is not merely a normalization factor; it is a compact representation of how the medium and the microscopic transport dynamics translate a given initial harmonic into the final momentum distribution. A comparison of $n=2$ and $n=3$ can therefore reveal whether the same nuclear-structure perturbation is propagated similarly in average-geometry dominated and fluctuation-dominated channels.

This is especially relevant for isobar collisions, which involve a comparison between systems containing the same number of nucleons, but different nuclear structure. In a Ru/Zr ratio, many global size effects are decreased and differences in deformation and proton-neutron density profiles remain. The ratio of Ru+Ru to Zr+Zr observables is then more differential than an absolute measurement in one of the two systems. The further ratio of a final-state flow ratio to an initial-state eccentricity ratio is even more selective, because it asks whether the leading initial-geometry difference is already sufficient to describe the observed final-state difference.

In addition to the motivation to detect the existence of anomalous electromagnetic effects, the RHIC isobar program (for $^{96}$Ru+$^{96}$Ru and $^{96}$Zr+$^{96}$Zr collisions at $\sqrt{s_{NN}}=200$ GeV) is a very precise tool for nuclear-structure studies \cite{STAR:2021mii,STAR:2024shape}. The two isobars differ in the values of the quadrupole and octupole deformation parameters and possibly in the proton-neutron density distribution. The quadrupole deformations for Ru are larger in commonly used structure inputs, while Zr has a large octupole deformation and large neutron skin. These differences impact not only the initial eccentricity distributions but also can be carried over to the final state anisotropic flow \cite{Li:2020vrg,Xu:2021qjw,Giacalone:2021clp,Zhang:2021kxj,Jia:2021wbq}.

The purpose of this work is to introduce and apply a geometry-normalized Ru/Zr response measure in AMPT. The existing isobar literature has made extensive use of separate Ru/Zr ratios of final flow and initial eccentricity. Here these two ingredients are combined into a single response diagnostic,
\begin{equation}
D_n(C)=
\frac{v_n^{\rm Ru}(C)/v_n^{\rm Zr}(C)}{\varepsilon_n^{\rm Ru}(C)/\varepsilon_n^{\rm Zr}(C)},
\label{eq:Dn_intro}
\end{equation}
where $C$ denotes a common $N_{\rm part}$ bin. We call $D_n$ an eccentricity-normalized isobar response double ratio.  It is intended to respond to a specific question: does the flow ratio after dividing out the leading Ru/Zr eccentricity ratio still have a residual harmonic dependent response?  The AMPT final states and the eccentricity records are read event by event and then averaged in common fixed-$N_{\rm part}$ intervals, so the comparison is performed for participant-matched ensembles rather than for unequal centrality-percentile classes. When the Ru/Zr difference in $v_n$ is completely determined by the Ru/Zr difference in $\varepsilon_n$ the value of $D_n$ equals unity. The deviations from unity are thus a measure of the portion (not eliminated by leading eccentricity scaling) of the isobar flow ratio.

Four configurations of AMPT, namely Ru+Ru deformation only, Zr+Zr deformation only, Ru+Ru deformation plus neutron skin and Zr+Zr deformation plus neutron skin, are analyzed. In all cases, the initial nuclei are deformed. The skin switch is off in samples with deformation only, and on in samples with deformation and skin. There are approximately $3.26\times 10^5$ AMPT events in each final-state sample. Eight systematic variations have been applied to the final production of the entire statistics, including different $N_{\rm part}$ binning, removing of peripheral bins, different $p_T$-ranges, and changes of the rapidity/pseudorapidity acceptance.

The main result is easy to state, and of physical interest. The elliptic-flow double ratio is approximately one, $D_2\simeq 1$ but the triangular-flow double ratio is consistently larger than one, $D_3>1$. The $D_3$ enhancement is present in both deformation-only and deformation-plus-skin, and is simulated with a stand-alone 2-subevent flow definition. The observation gives a model-resolved statement about AMPT response: Ru/Zr elliptic-flow differences are largely controlled by Ru/Zr ellipticity differences, while triangular flow retains an additional response component after the triangularity ratio is divided out.

The paper is organized as follows. Section~\ref{sec:model} describes the nuclear inputs and the AMPT/HIJING source modification. Section~\ref{sec:observables} defines the eccentricities, flow coefficients, response ratios and uncertainty treatment. Section~\ref{sec:results} presents the full-statistics results. Section~\ref{sec:systematics} discusses systematic variations and validation checks. Section~\ref{sec:discussion} gives the physical interpretation. Section~\ref{sec:conclusions} summarizes the conclusions.

\section{AMPT setup and nuclear-structure inputs}
\label{sec:model}

\subsection{AMPT and the modified HIJING initialization}

The calculations are made using the string-melting version of the AMPT model, which uses fluctuating initial conditions from HIJING, parton scattering in ZPC, quark coalescence and hadronic rescattering \cite{Wang:1991hta,Gyulassy:1994ew,Zhang:1999bd,Lin:2004en,Lin:2014tya}. AMPT is not a first-principles hydrodynamic calculation, but rather a microscopic transport model with finite scattering and coalescence dynamics. For the present question this is helpful since the same generator, transport parameters, and final state analysis can be used for all four nuclear configurations. The comparison is thus focused on the controlled change of the input nuclear density and not on changes of the following analysis chain.

The HIJING initialization used here was modified to sample deformed Woods--Saxon nuclei. For a deformed density the radial scale is written as
\begin{equation}
R(\theta)=R_0\left[1+\beta_2Y_{20}(\theta)+\beta_3Y_{30}(\theta)+\beta_4Y_{40}(\theta)+\beta_6Y_{60}(\theta)\right],
\label{eq:deformed_radius}
\end{equation}
and the density is sampled from a Woods--Saxon profile,
\begin{equation}
\rho(r,\theta)=\frac{\rho_0}{1+\exp\left[(r-R(\theta))/a\right]}.
\label{eq:woods_saxon}
\end{equation}
In this case, $R_0$ is the nuclear radius, $a$ is the diffuseness, $Y_{\ell0}$ are real spherical harmonics in the intrinsic frame of the nucleus, and $\beta_{\ell}$ are deformation coefficients. For the present production, the quadrupole and, for Zr, the octupole components related to the Ru/Zr isobar problem are the active deformations. The higher symbols in eq.~\eqref{eq:deformed_radius} are kept for describing the source structure but do not make any further claims in the analysis. The production runs are carried out with random nuclear orientations.

In the deformation-plus-skin samples, the Woods--Saxon parameters for proton and neutron are taken separately. In the modified HIJING source, the flag for the neutron-skin is toggled between the common density and the separate proton/neutron densities. If skin is on, the proton and neutron coordinates are calculated using their own radii and diffuseness parameters. If it is off, a single distorted density is applied, with the same deformation parameters. The difference is, then, not between deformed and spherical nuclei, but between deformed nuclei with no explicit proton-neutron density separation and deformed nuclei with an explicit proton-neutron density separation. The source also writes the eccentricity record to be used by the analysis, which enables the final-state flow observables to be compared with the initial-state flow observables in the same interval of the number of participants.

\subsection{Four nuclear configurations}

The four systems considered are: $^{96}$Ru+$^{96}$Ru (deformation only), $^{96}$Zr+$^{96}$Zr (deformation only), $^{96}$Ru+$^{96}$Ru (deformation and neutron skin), and $^{96}$Zr+$^{96}$Zr (deformation and neutron skin). The parameters describing the deformation and neutron-skin of the isobars are taken from the set of Woods-Saxon parameters discussed in the isobar nuclear-structure analysis of Nijs and van der Schee \cite{Nijs:2023isobar}, given in the table \ref{tab:ws} here. That work presents separate proton and neutron Woods-Saxon radii and diffuseness parameters for isobars, and highlights case 5, which includes the proton-neutron density separation and deformation inputs based on the STAR isobar measurements. We use the corresponding case-5 values for the skin-included samples.

The deformation-only samples retain the same $\beta_2$ and $\beta_3$ values but use a common proton-neutron density. This construction allows for a controlled reference with the deformation content, but without the extra neutron-skin separation. The label ``deformation only'' is important in the interpretation since it does not imply that the nuclei are spherical. It implies that the explicit proton/neutron density splitting is not turned on.

\begin{table}[t]
\centering
\caption{Nuclear-structure input for the four AMPT configurations. The skin-included Ru and Zr values are case-5 Woods--Saxon parameters as listed in ref.~\cite{Nijs:2023isobar}. For the deformation-only samples, the same deformation parameters are retained while the proton-neutron skin separation is switched off. The deformation-only rows use common proton-neutron densities with the same deformation parameters, while the skin-included rows use separate proton and neutron densities.}
\label{tab:ws}
\resizebox{\textwidth}{!}{%
\begin{tabular}{llcccccc}
\toprule
Nucleus & configuration & $R_p$ [fm] & $a_p$ [fm] & $R_n$ [fm] & $a_n$ [fm] & $\beta_2$ & $\beta_3$ \\
\midrule
$^{96}$Ru & skin included & $5.053$ & $0.480$ & $5.073$ & $0.490$ & $0.154$ & $0$ \\
$^{96}$Zr & skin included & $4.912$ & $0.508$ & $5.007$ & $0.564$ & $0.062$ & $0.202$ \\
$^{96}$Ru & deformation only & $5.065$ & $0.485$ & $5.065$ & $0.485$ & $0.154$ & $0$ \\
$^{96}$Zr & deformation only & $4.961$ & $0.544$ & $4.961$ & $0.544$ & $0.062$ & $0.202$ \\
\bottomrule
\end{tabular}}
\end{table}

The difference between deformation only and deformation plus skin is of importance. The deformation-only sample is not a spherical reference sample. It includes the quadrupole and, if applicable, octupole deformation, but there are no separate proton and neutron density profiles. The deformation-plus-skin sample has both the same deformation parameters and the proton-neutron density separation. The paper thus compares two physics cases for each pair of isobars: deformation-only and deformation-plus-skin. The same initialisation block is used for projectile and target nuclei in the source implementation. If the switch is turned off for the neutron-skin, then the parameters of common density are used. If the switch is turned on, the proton and neutron diffuseness and radii parameters are used independently.

\begin{figure}[H]
\centering
\includegraphics[width=1.0\textwidth]{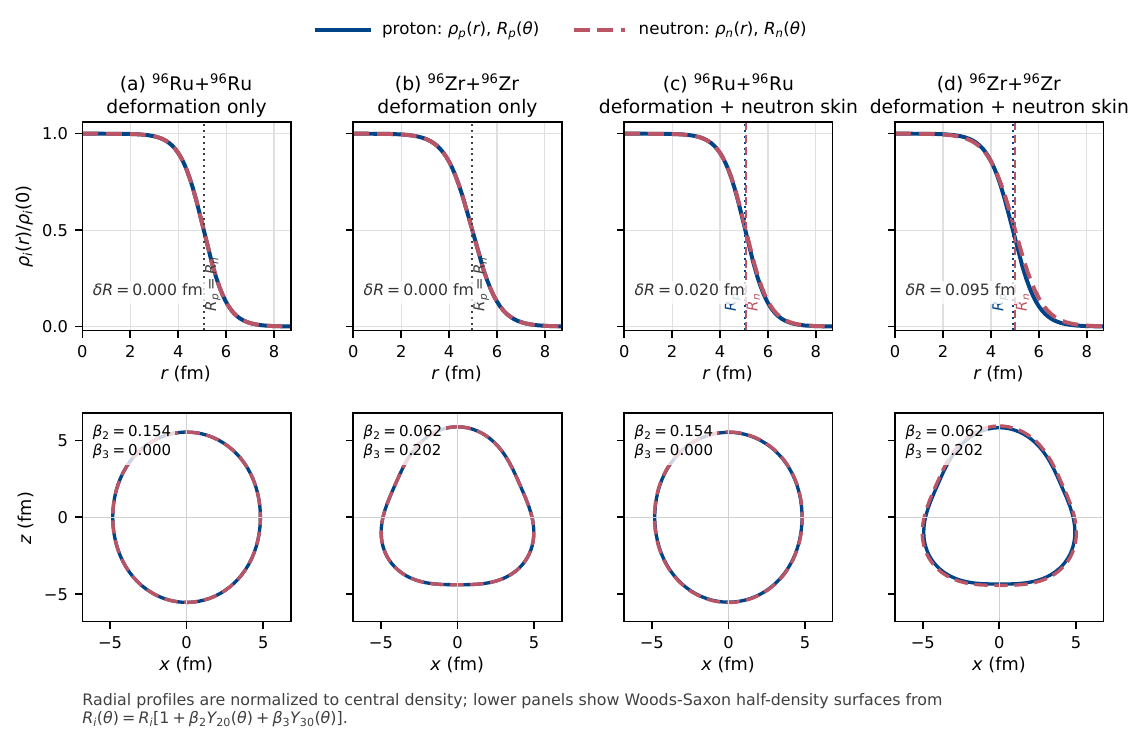}
\caption{The four AMPT configurations with proton and neutron Woods–Saxon density inputs. Normalized radial density profiles are plotted in the upper panels and half-density surfaces in the intrinsic frame in the lower panels. Proton and neutron profiles are assumed to be the same in the deformation-only samples, and different in the deformation-plus-skin samples. The labels indicate the deformation parameters and the displacement of the neutron skin $\delta R=R_n-R_p$.}
\label{fig:nuclear_inputs}
\end{figure}

The density inputs are represented compactly in a visual way in figure~\ref{fig:nuclear_inputs}. The radial profiles show the proton-neutron separation that is not present in the deformation-only samples, but is present in the skin-included samples. The bottom panels indicate that the shape information is not restricted to a radial change; the quadrupole deformation of Ru and the octupole component of Zr remain in both physics cases. This figure is intended as an input-geometry guide for the reader; the quantitative analysis below uses the event-by-event eccentricity records generated from these density inputs.

\subsection{Event sample and analysis cuts}

The total number of AMPT events for the full-statistics production is $325582$ for each one of the four collision configurations. The final-state charged particle flow analysis is performed in the range $0.2<p_T<5.0$ GeV/$c$. The selection of the nominal final-state particle uses $|y|<0.5$ and the event-plane reference subevents use $|\eta|<0.5$. These cuts are deliberately made symmetric and simple between Ru and Zr. They are not optimized individually for each system, as this would add an artificial response difference. Systematic study repeats the analysis using the same analysis definition but with different values for the interval of the transverse momentum and the rapidity/pseudorapidity acceptance, and checks the stability of the double ratios without changing the central analysis definition.

All the observables are calculated in the same fixed-$N_{\rm part}$ bins. The nominal six intervals are $108\leq N_{\rm part}<189$, $60\leq N_{\rm part}<108$, $30\leq N_{\rm part}<60$, $13\leq N_{\rm part}<30$, $4\leq N_{\rm part}<13$ and $2\leq N_{\rm part}<4$. Binning by participant number is an integral part of the analysis. It does not involve comparing Ru and Zr events which happen to occur in different percentile intervals of the multiplicity distribution, but rather makes the double ratio a direct test of the geometry-to-flow transfer at a comparable size of overlap. Accepted publication summaries are only created and used with the bins that meet the pre-defined quality requirements. If values fail these requirements, they are not deleted from the record, but are left in the diagnostic outputs to be examined to find the source of each acceptance decision.

\section{Observables and analysis method}
\label{sec:observables}

\subsection{Initial eccentricities}

The transverse density distribution is characterized by participant eccentricities. The complex eccentricity is represented by
\begin{equation}
\varepsilon_n e^{in\Phi_n}
= -\frac{\int r^n e^{in\phi}\rho(r,\phi)r\,dr\,d\phi}{\int r^n\rho(r,\phi)r\,dr\,d\phi}.
\label{eq:eccentricity}
\end{equation}
In the above formula, $\Phi_n$ is the angle of the participant plane \cite{Bhalerao:2011yg,Niemi:2012aj,ATLAS:2013xzf}. Eccentricity files generated by the present AMPT have 18 columns: the first two are for $\varepsilon_2$ and $\varepsilon_3$. They do not have a specific column for the number $\varepsilon_4$. Hence, only nonlinear diagnostics are reported for the 4th order and not directly as response coefficients $\kappa_4$ and $D_4$.

For each system $A$ and fixed-$N_{\rm part}$ bin $C$, the analysis computes the mean eccentricity $\langle\varepsilon_n\rangle_A(C)$ and the RMS eccentricity $\varepsilon_{n,{\rm rms}}^A(C)=\sqrt{\langle\varepsilon_n^2\rangle_A(C)}$. The mean eccentricity is used for the nominal double ratios because it matches the bin-wise response definition used for the integrated event-plane flow. The RMS eccentricity is kept as a useful cross-check as RMS quantities are more similar to two-particle cumulant observables and are less influenced by sign conventions in the participant plane \cite{ALICE:2011ab,CMS:2013bza}. If the major trend is the same in both the mean and RMS eccentricity analyses, then it is more likely that it is a true result and not due to a special eccentricity averaging prescription.

The uncertainties in eccentricity are computed in the same binning structure as the final-state flow. This is important as the final and initial quantities are added together bin by bin in this double ratio. A small Ru/Zr mismatch in the mean participant number can mimic a response difference if the underlying eccentricity or flow changes rapidly with centrality. For this reason the analysis records the mean $N_{\rm part}$ for every system and bin and requires the accepted bins to satisfy a participant-matching quality condition.

\subsection{Charged-particle flow}

The azimuthal distribution of final-state charged particles is expanded as
\begin{equation}
\frac{dN}{d\phi}\propto 1+2\sum_{n=1}^{\infty}v_n\cos[n(\phi-\Psi_n)] .
\label{eq:flow_expansion}
\end{equation}
The two-subevent event-plane method is used for the nominal integrated flow \cite{Poskanzer:1998yz,Voloshin:2008dg}. Negative- and positive-pseudorapidity reference subevents are constructed, their event-plane correlation gives a resolution correction, and particles in one side are correlated with the opposite-side plane. This reduces the short-range autocorrelation in comparison to a same-particle estimate of the event plane, and provides a direct measurement of the event-plane resolution. The same particle cuts and resolution requirements are applied to Ru and Zr, so a difference in $v_n$ cannot arise from a different analysis prescription in the two isobars. The analysis also calculates two-subevent two-particle cumulant-like values, $v_n\{2,{\rm sub}\}$, as a cross check \cite{Bilandzic:2010jr,Bilandzic:2013kga,Jia:2017hbm}.

The response coefficient in a given system and bin is defined as
\begin{equation}
\kappa_n^A(C)=\frac{v_n^A(C)}{\langle\varepsilon_n\rangle_A(C)}.
\label{eq:kappa}
\end{equation}
This coefficient is model and cut dependent. It does not signify a medium property that is common to all media. It is given to compare the conversion of Ru and Zr eccentricities to final state flow in identical AMPT setups. The response coefficient is useful especially in isobar mode, as many of the common effects in AMPT, such as the global transport setup and hadronic afterburning treatment, are common to both modes. A residual Ru/Zr difference in $\kappa_n$ is therefore a more selective diagnostic than the absolute magnitude of $v_n$ alone.
\subsection{Ru/Zr ratios and double ratios}

For each case, deformation-only or deformation-plus-skin, the Ru/Zr eccentricity and flow ratios are
\begin{equation}
R_{\varepsilon_n}(C)=\frac{\varepsilon_n^{\rm Ru}(C)}{\varepsilon_n^{\rm Zr}(C)},
\qquad
R_{v_n}(C)=\frac{v_n^{\rm Ru}(C)}{v_n^{\rm Zr}(C)}.
\label{eq:ratios}
\end{equation}
The double ratio is then
\begin{equation}
D_n(C)=\frac{R_{v_n}(C)}{R_{\varepsilon_n}(C)}
=\frac{v_n^{\rm Ru}(C)/v_n^{\rm Zr}(C)}{\varepsilon_n^{\rm Ru}(C)/\varepsilon_n^{\rm Zr}(C)}.
\label{eq:double_ratio}
\end{equation}
Equivalently, $D_n$ compares the Ru and Zr response coefficients. If $v_n/\varepsilon_n$ is the same in the two isobars, then $D_n=1$.

The accepted-bin integrated value is computed as an inverse-variance weighted mean over bins passing the publication-quality selections. The analysis block-resampling procedure is used to get statistical uncertainties. The systematic uncertainty is derived by performing the analysis again with the pre-defined variations. The standard systematic spread is the standard deviation of the variation set, while a systematic envelope is the maximum absolute deviation from the nominal value. The total standard uncertainty and the envelope total uncertainty are
\begin{equation}
\sigma_{\rm tot,std}=\sqrt{\sigma_{\rm stat}^2+\sigma_{\rm syst,std}^2},
\qquad
\sigma_{\rm tot,env}=\sqrt{\sigma_{\rm stat}^2+\sigma_{\rm syst,env}^2} .
\label{eq:total_unc}
\end{equation}
In this case, $\sigma_{\rm stat}$ is the nominal statistical uncertainty, $\sigma_{\rm syst,std}$ is the standard deviation of the systematic variations, and $\sigma_{\rm syst,env}$ is the maximum absolute value of the systematic displacement from the nominal value.

\subsection{Quality selections and validation tests}

All raw bins are saved during the analysis, but the bins are only marked as accepted after the quality cuts on AMPT event count, eccentricity event count, charged-particle flow count, event-plane resolution, denominator stability and relative uncertainty. These cuts are not designed to achieve a desired physics conclusion, they are designed to exclude points that are peripheral or have lower resolution or denominator.

Multiple internal validity checks are performed. The code verifies that the most central bin index is not lost when reading tabulated analysis summaries; this is important because bin index zero must remain zero. It also checks that the shifted observables $D_n-1$ are compared to reference value 0, while the unshifted observables $D_n$ are compared to reference value 1. The code creates accepted-point lists, systematic-run summaries and integrity summaries. Null tests include split-half same-system ratios and Ru/Zr label shuffling. These tests are used as diagnostics to see if the accepted double-ratio structure is stable to purely statistical splits.

\section{Full-statistics results}
\label{sec:results}

\subsection{Initial eccentricities}

\begin{figure}[t]
\centering
\includegraphics[width=1.0\textwidth]{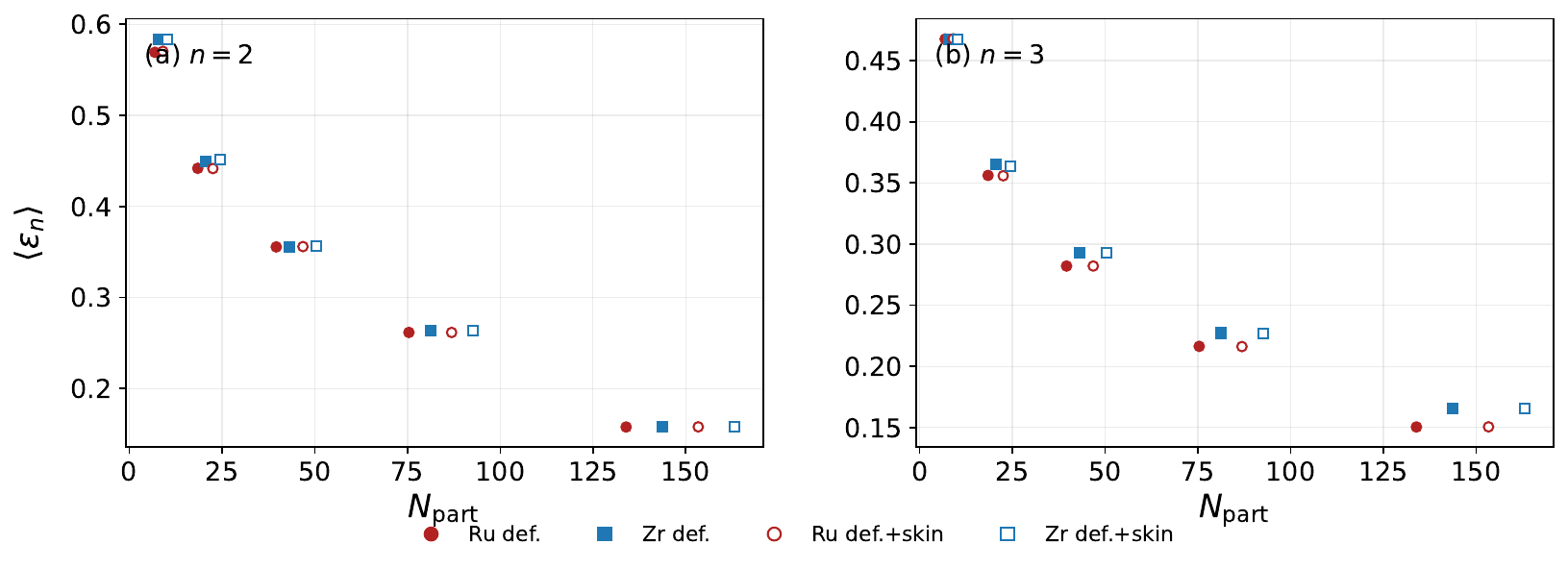}
\caption{The mean eccentricities $\langle\varepsilon_2\rangle$ and $\langle\varepsilon_3\rangle$ for the participants as functions of $N_{\rm part}$ in deformation-only and deformation-plus-skin configurations for Ru+Ru and Zr+Zr. Common fixed-$N_{\rm part}$ bins are used for the analysis, where small horizontal offsets are only provided for the sake of clarity.}
\label{fig:eccentricities}
\end{figure}

The mean eccentricities $\langle\varepsilon_2\rangle$ and $\langle\varepsilon_3\rangle$ are plotted as a function of $N_{\rm part}$ in figure~\ref{fig:eccentricities}. The common-bin structure is evident; all four systems are measured during the same intervals of participant numbers, and small horizontal offsets are only provided for the sake of clarity. The ellipticity is a measure of the deformation and average geometry of the isobar pair. The triangularity is more sensitive to fluctuation and to the octupole deformed Zr input.

\begin{figure}[t]
\centering
\includegraphics[width=1.0\textwidth]{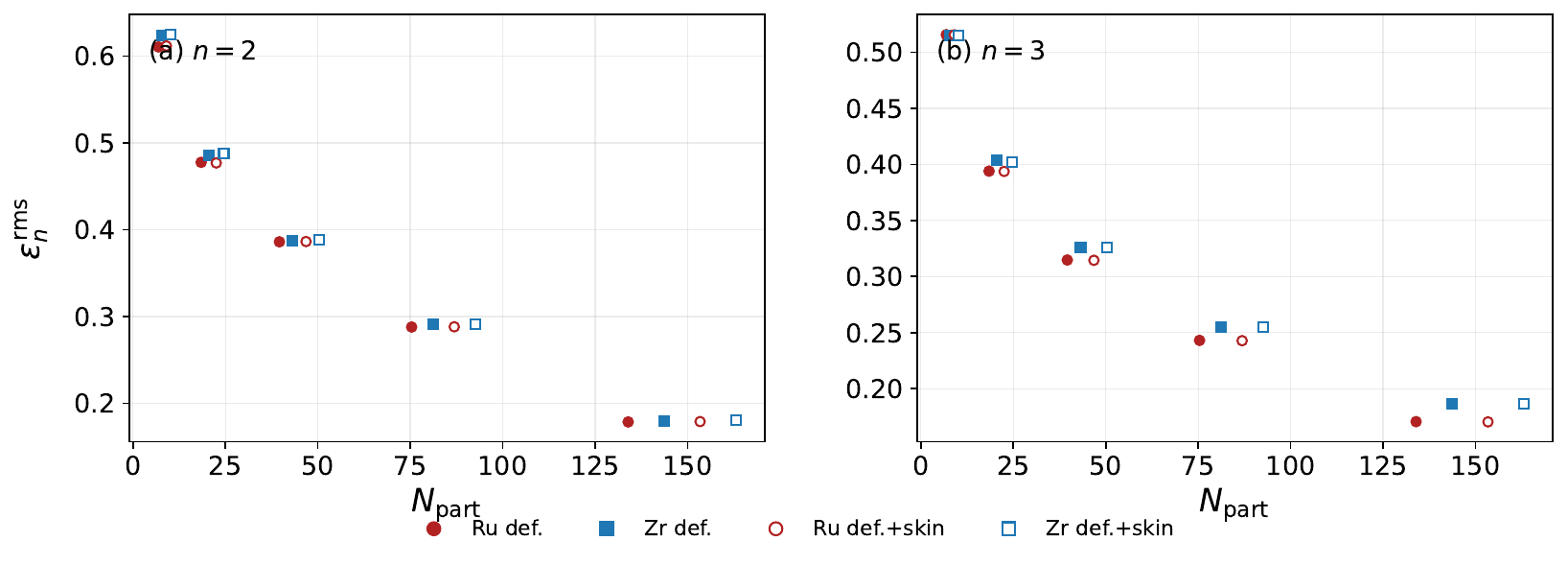}
\caption{RMS eccentricities $\varepsilon_{2,{\rm rms}}$ and $\varepsilon_{3,{\rm rms}}$ in the same fixed-$N_{\rm part}$ intervals. RMS eccentricities are retained as a cross-check of the mean-eccentricity double ratios.}
\label{fig:eccentricities_rms}
\end{figure}

The RMS eccentricities can be plotted as in figure~\ref{fig:eccentricities_rms}. These values are useful since two-particle observables may be more sensitive to RMS eccentricities than to mean eccentricities. The qualitative Ru/Zr ordering is stable when the mean eccentricity is replaced by the RMS eccentricity. It is crucial for the interpretation of the double ratios since it demonstrates that the main hierarchy is not due to a special averaging of the eccentricity.

\subsection{Charged-particle flow}

\begin{figure}[t]
\centering
\includegraphics[width=1.0\textwidth]{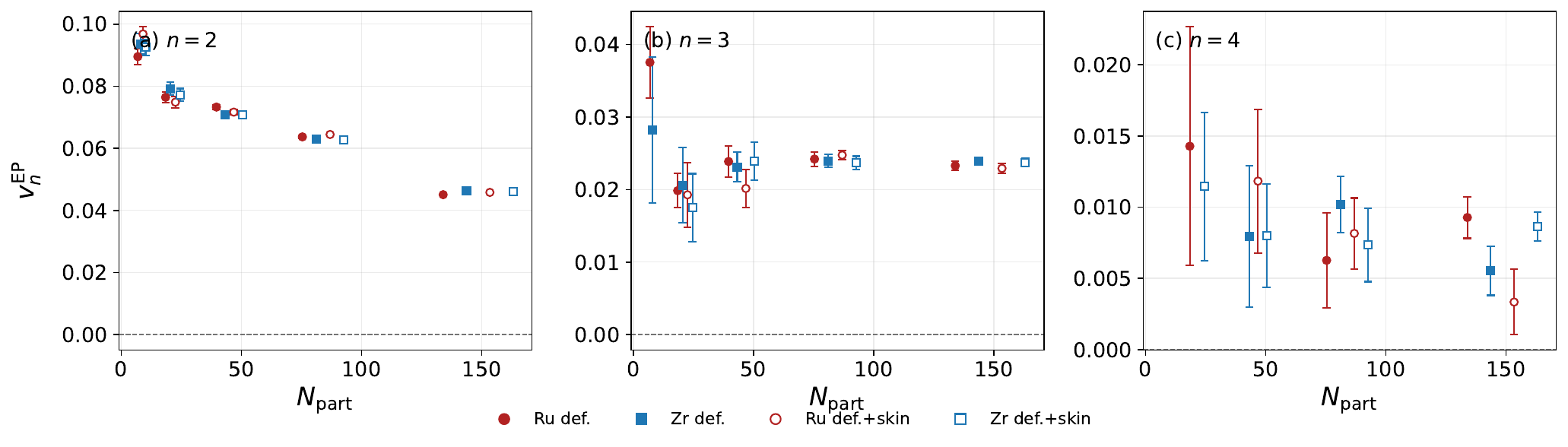}
\caption{Integrated charged-particle $v_2$ and $v_3$ as a function of $N_{\rm part}$ for the four AMPT configurations. The nominal event-plane values are used for the primary response coefficients. The two-subevent check is shown separately
through the double-ratio and integrated-summary results.}
\label{fig:flow}
\end{figure}

Figure~\ref{fig:flow} presents the charged-particle $v_2$ and $v_3$ obtained from the nominal event-plane analysis. As expected, the elliptic flow is greater and more accurate than the triangular flow. The triangular-flow points have higher relative uncertainties, particularly at low $N_{\rm part}$, due to the smaller event-plane resolution at the third harmonic, and the increased sensitivity to multiplicity.

\begin{figure}[t]
\centering
\includegraphics[width=1.0\textwidth]{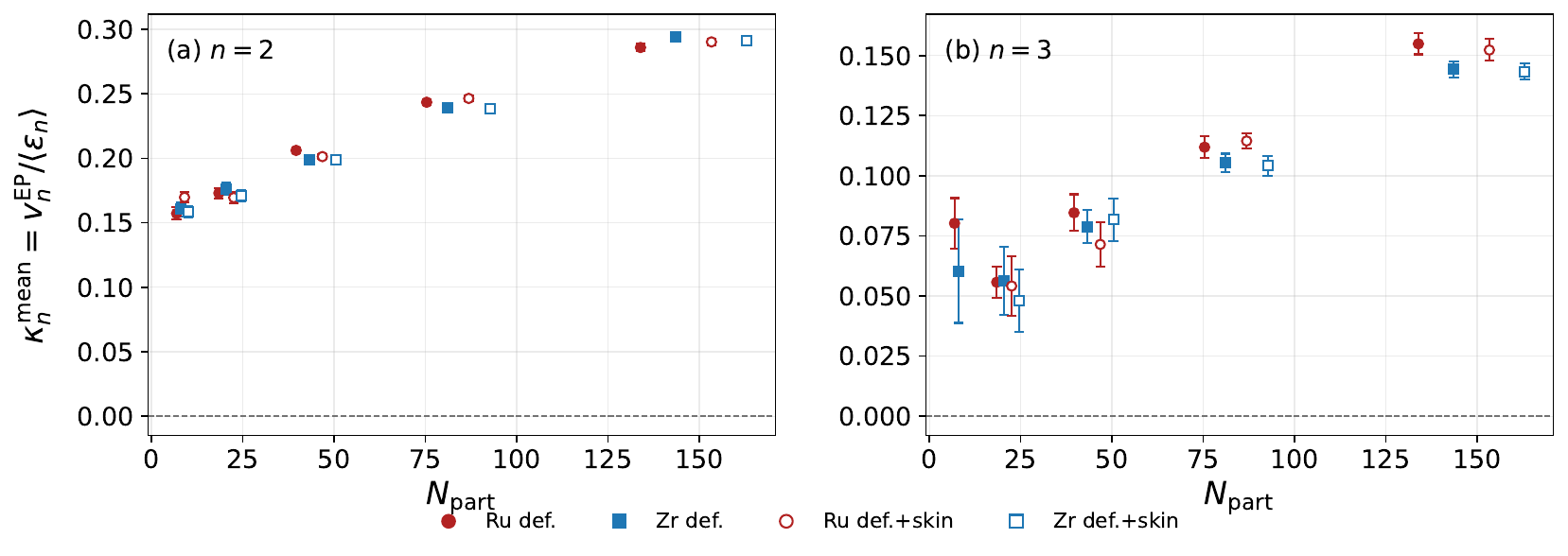}
\caption{The linear response-coefficients $\kappa_2$ and $\kappa_3$ for the four systems. These are effective AMPT response measures which have been assessed in the same cuts and fixed-$N_{\rm part}$ bins for Ru and Zr.}
\label{fig:kappa}
\end{figure}

The response coefficients $\kappa_n=v_n/\langle\varepsilon_n\rangle$ are plotted in figure~\ref{fig:kappa} and tabulated in table~\ref{tab:kappa}. The response coefficient gets smaller towards the outer bins due to lower multiplicity, less collectivity and worse event-plane resolution for small systems. The absolute values of $\kappa_n$ is not the most important result as it is dependent on the settings of the AMPT and the analysis cuts. What is important is the relative Ru/Zr behaviour at a constant $N_{\rm part}$.

\begin{table}[t]
\centering
\caption{Accepted-bin weighted response coefficients for nominal full-statistics run. The uncertainties are statistical/block-resampling uncertainties related to the nominal sample.}
\label{tab:kappa}
\small
\begin{tabular}{llcccc}
\toprule
Nucleus & configuration & harmonic & $\kappa_n$ & stat. unc. & accepted bins \\
\midrule
Ru & deformation only & 2 & $0.2244$ & $0.0013$ & 5 \\
Ru & deformation + skin & 2 & $0.2282$ & $0.0012$ & 5 \\
Zr & deformation only & 2 & $0.2328$ & $0.0015$ & 5 \\
Zr & deformation + skin & 2 & $0.2364$ & $0.0013$ & 5 \\
Ru & deformation only & 3 & $0.1134$ & $0.0026$ & 5 \\
Ru & deformation + skin & 3 & $0.1201$ & $0.0024$ & 4 \\
Zr & deformation only & 3 & $0.1200$ & $0.0023$ & 5 \\
Zr & deformation + skin & 3 & $0.1207$ & $0.0025$ & 4 \\
\bottomrule
\end{tabular}
\end{table}

\subsection{Eccentricity ratios, flow ratios and double ratios}

\begin{figure}[t]
\centering
\includegraphics[width=1.0\textwidth]{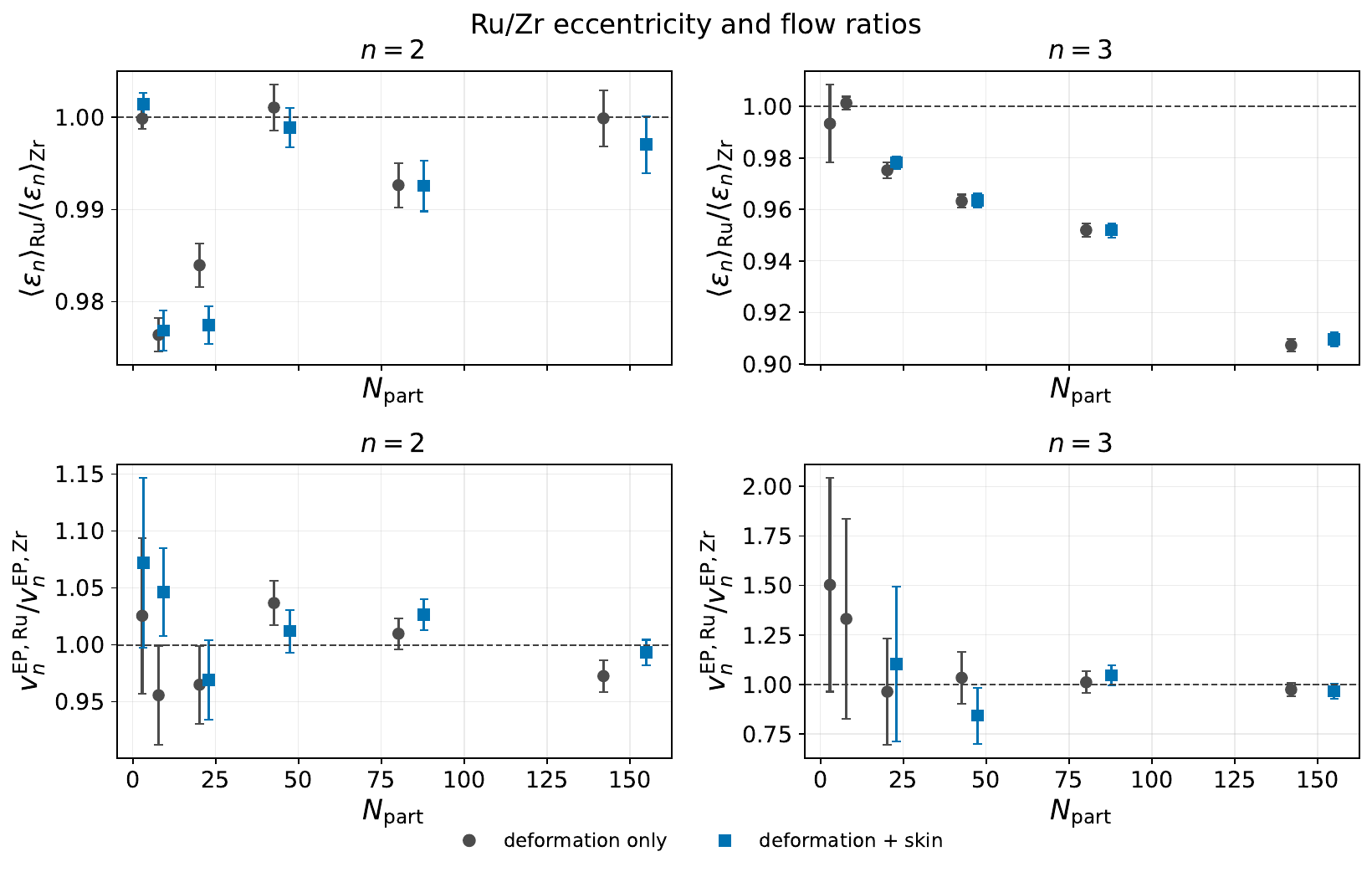}
\caption{Isolate Ru/Zr eccentricity ratios and charged-flow ratios. The double ratio $D_n$ is obtained when the flow ratio is divided by the eccentricity ratio in a fixed-$N_{\rm part}$ bin.}
\label{fig:ratios}
\end{figure}

Figure~\ref{fig:ratios} separates the ingredients of the double ratio. It calculates and displays the eccentricity ratios (Ru/Zr) and flow ratios (Ru/Zr) before the geometry will be normalized with the final geometry. This separation may be helpful in determining if a final double ratio is not equal to one due to initial geometry, final response or both. The double-ratio construction is not model independent, but it eliminates the leading Ru/Zr eccentricity ratio, and so separates out any remaining differences in the response.

\begin{figure}[t]
\centering
\includegraphics[width=0.96\textwidth]{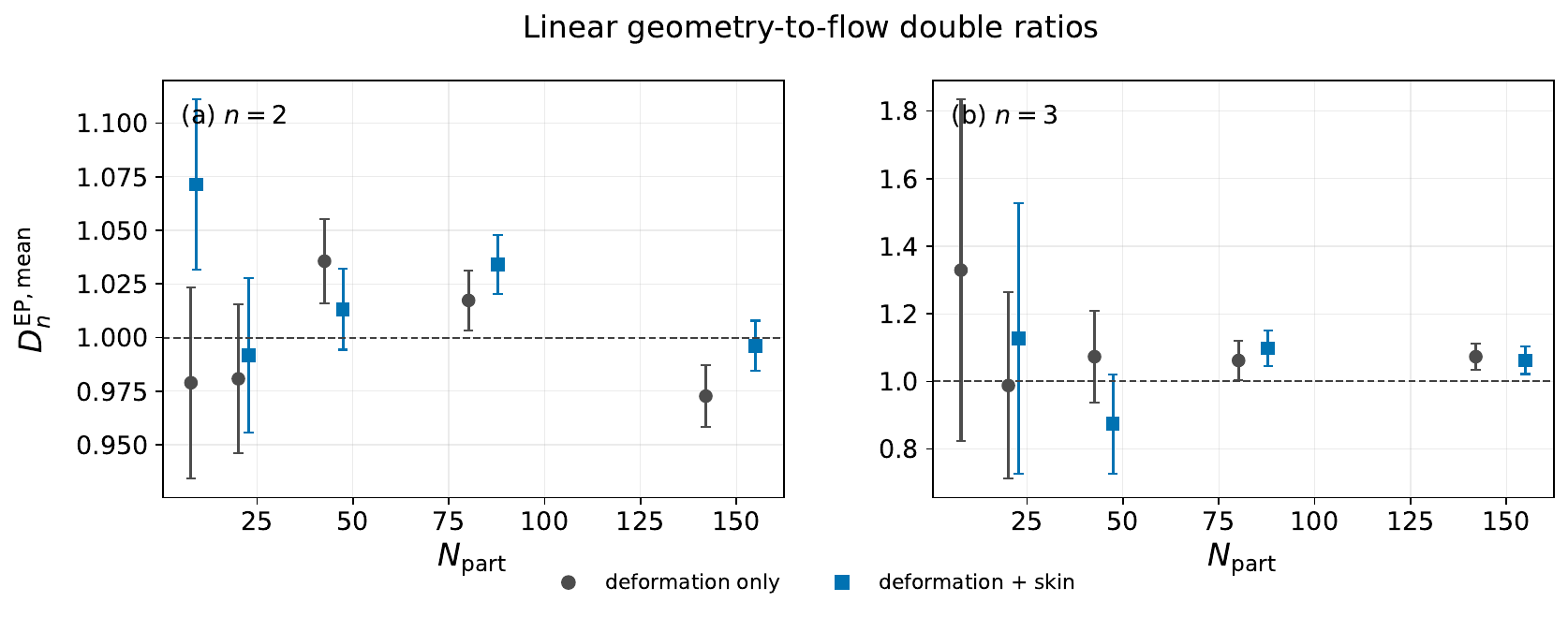}
\caption{Geometry-normalized double ratios $D_2$ and $D_3$ versus $N_{\rm part}$. The horizontal line represents $D_n=1$. The elliptic double ratio is consistent with unity, and the triangular double ratio is consistently above unity in the accepted bins.}
\label{fig:double_ratios}
\end{figure}

Figure~\ref{fig:double_ratios} shows the main bin-wise observable $D_n$. The horizontal line represents $D_n=1$ at pure eccentricity scaling using the common Ru/Zr response coefficient. The two harmonics have different characteristics. The accepted values for $n=2$ oscillate around the value 1. For $n=3$, the accepted values are mostly above unity with some higher uncertainty. The key finding of the analysis is the separation between the two harmonics. It demonstrates that the double-ratio construction is not simply mimicking a normalization shift, but instead it offers a particular solution to a mismatch for the elliptic and triangular response channels.

The integrated double-ratio summary including systematic uncertainties is given in table~\ref{tab:final}. The systematic envelope uncertainties are also reported in the same table. The triangular-flow deviations for the systematic envelope are approximately $1.7$--$1.8$ standard deviations. What is important is the sign and the stability of the hierarchy: all four $D_3$ summaries are greater than unity, while the $D_2$ summaries are close to unity. The result is thus best represented as a quantitative bounded, internally stable AMPT response pattern.

\begin{table}[t]
\centering
\caption{Double-ratio summary in accepted-bins, from systematic run with full-statistics. The statistical uncertainty is the nominal block-resampling uncertainty. The systematic standard deviation is obtained from the variation set. The envelope is the maximum absolute deviation from the nominal result. $Z_{\rm std}$ and $Z_{\rm env}$ are the significances relative to $D_n=1$ using the standard and envelope total uncertainties, respectively.}
\label{tab:final}
\small
\resizebox{\textwidth}{!}{%
\begin{tabular}{llcccccccc}
\toprule
Observable & configuration & nominal $\pm$ stat. & $\sigma_{\rm syst,std}$ & $\sigma_{\rm syst,env}$ & $\sigma_{\rm tot,std}$ & $\sigma_{\rm tot,env}$ & $Z_{\rm std}$ & $Z_{\rm env}$ & bins \\
\midrule
$D_2^{\rm EP}$ & deformation only & $1.0016\pm0.0085$ & $0.0022$ & $0.0040$ & $0.0087$ & $0.0093$ & $0.18$ & $0.17$ & 5 \\
$D_2^{\rm EP}$ & deformation + skin & $1.0134\pm0.0077$ & $0.0031$ & $0.0086$ & $0.0083$ & $0.0116$ & $1.61$ & $1.16$ & 5 \\
$D_2\{2,{\rm sub}\}$ & deformation only & $1.0051\pm0.0075$ & $0.0022$ & $0.0046$ & $0.0078$ & $0.0088$ & $0.66$ & $0.59$ & 5 \\
$D_2\{2,{\rm sub}\}$ & deformation + skin & $1.0116\pm0.0069$ & $0.0027$ & $0.0075$ & $0.0075$ & $0.0102$ & $1.55$ & $1.13$ & 5 \\
$D_3^{\rm EP}$ & deformation only & $1.0699\pm0.0312$ & $0.0089$ & $0.0216$ & $0.0325$ & $0.0379$ & $2.15$ & $1.84$ & 5 \\
$D_3^{\rm EP}$ & deformation + skin & $1.0672\pm0.0312$ & $0.0086$ & $0.0236$ & $0.0324$ & $0.0391$ & $2.08$ & $1.72$ & 4 \\
$D_3\{2,{\rm sub}\}$ & deformation only & $1.0825\pm0.0307$ & $0.0155$ & $0.0355$ & $0.0344$ & $0.0469$ & $2.40$ & $1.76$ & 5 \\
$D_3\{2,{\rm sub}\}$ & deformation + skin & $1.0921\pm0.0341$ & $0.0152$ & $0.0396$ & $0.0373$ & $0.0522$ & $2.47$ & $1.76$ & 3 \\
\bottomrule
\end{tabular}}
\end{table}

\section{Systematic variations and validation}
\label{sec:systematics}

\begin{figure}[t]
\centering
\includegraphics[width=1.0\textwidth]{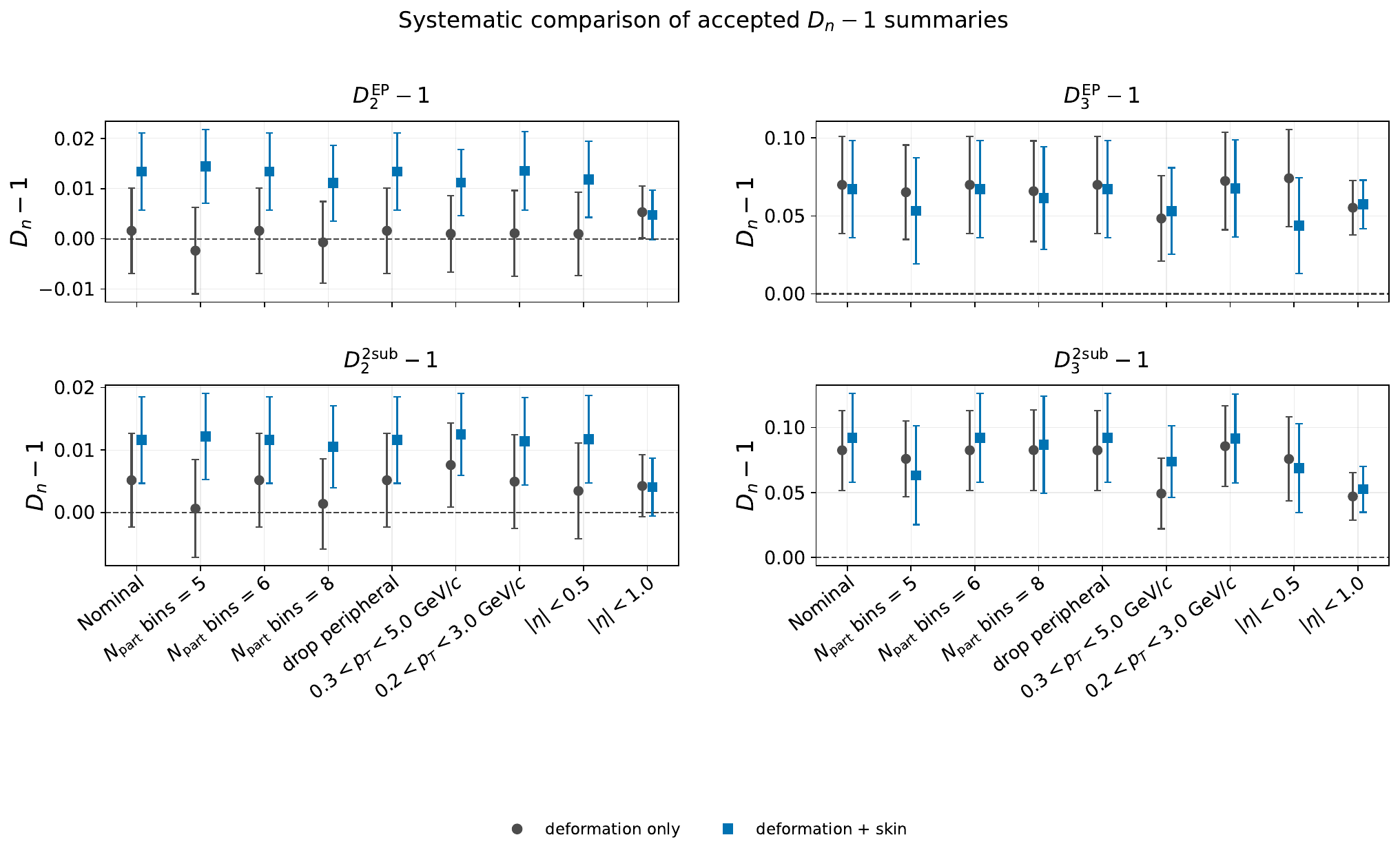}
\caption{Systematic comparison of accepted value of integrated $D_n-1$. The horizontal line marks zero, corresponding to $D_n=1$. The $D_3$ observables are positive for the tested variations, and the $D_2$ observables are very close to zero.}
\label{fig:systematics}
\end{figure}

The final systematic study is based on the variation set in table~\ref{tab:systematics}. All variations successfully finished in full production. The systematic comparison is shown in figure~\ref{fig:systematics}. The figure plots $D_n-1$ such that the reference line is at zero. This representation highlights the hierarchy in particular: $D_2-1$ is consistent with zero within uncertainties while $D_3-1$ is positive over the accepted systematic set.

The systematic variations are intended to investigate analysis options, not to retune the model. The number of $N_{\rm part}$ bins is changed to test the integrated result for the dependence on the specific bin boundaries. Dropping the most peripheral accepted bin tests sensitivity to the region where multiplicity and event-plane resolution are weakest. The $p_T$ variations test whether the response hierarchy is driven by the low- or high-transverse-momentum part of the charged-particle sample. Variations in the rapidity and pseudorapidity test whether the result depends on the final-state acceptance definition. This set of changes has preserved the same harmonic hierarchy, which is a major factor in the robustness of the result.

\begin{table}[t]
\centering
\caption{Systematic variations used in the full-statistics production run.}
\label{tab:systematics}
\small
\begin{tabularx}{\textwidth}{p{0.22\textwidth}p{0.50\textwidth}p{0.20\textwidth}}
\toprule
Variation & definition & purpose \\
\midrule
Nominal & six common quantile $N_{\rm part}$ bins, $0.2<p_T<5.0$ GeV/$c$, $|y|<0.5$ & central result \\
$N_{\rm part}$ bins = 5 & repeat with five common bins & binning dependence \\
$N_{\rm part}$ bins = 8 & repeat with eight common bins & binning dependence \\
Drop peripheral & remove the lowest-$N_{\rm part}$ bin from accepted summaries & peripheral stability \\
$0.3<p_T<5.0$ GeV/$c$ & raise lower flow cut & low-$p_T$ sensitivity \\
$0.2<p_T<3.0$ GeV/$c$ & lower upper flow cut & high-$p_T$ sensitivity \\
$|\eta|<0.5$ & use pseudorapidity selection & acceptance definition \\
$|\eta|<1.0$ & widen pseudorapidity selection & acceptance size \\
\bottomrule
\end{tabularx}
\end{table}

\begin{figure}[t]
\centering
\includegraphics[width=0.90\textwidth]{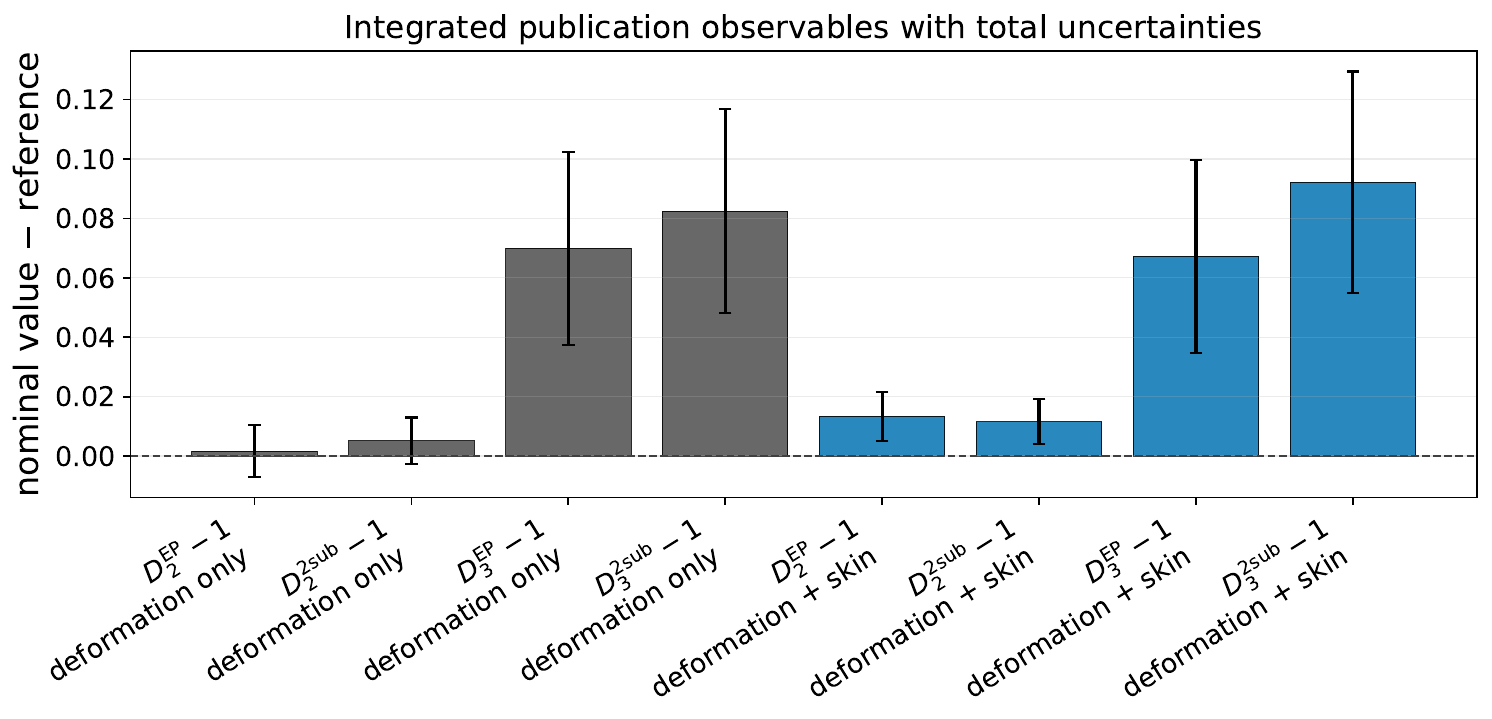}
\caption{Final integrated summary of $D_2-1$ and $D_3-1$ with total uncertainties. The triangular double ratios are positive in both nuclear configurations and in both flow definitions, whereas the elliptic double ratios are consistent with zero deviation from unity.}
\label{fig:integrated}
\end{figure}

A compact final integrated summary with total uncertainties is presented in figure~\ref{fig:integrated}. Combined with table~\ref{tab:final}, this figure provides the best overall picture of the result. The central production estimate is represented by the standard systematic total uncertainty, and the envelope uncertainty represents the range of the variations of the tested analysis.

\begin{figure}[t]
\centering
\includegraphics[width=1.0\textwidth]{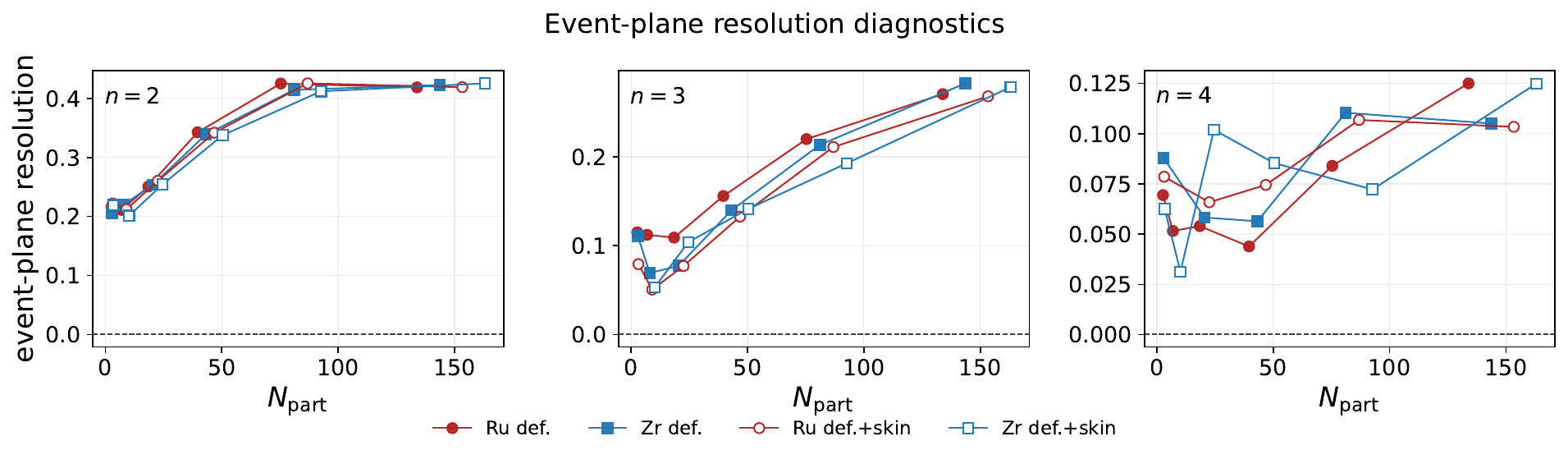}
\caption{Event-plane resolution diagnostics for the nominal full-statistics run. Low-$N_{part}$ bins are the most vulnerable to poor resolution and are removed by the accepted-point quality selections.}
\label{fig:resolution}
\end{figure}

The event-plane resolution diagnostics are shown in figure~\ref{fig:resolution}. Expected behaviour is seen: resolution is worse towards the periphery and decreases with higher harmonics. The accepted point selection eliminates bins for which the response ratios would not be considered stable due to resolution or denominator instability. This is the reason why some $D_3$ deformation-plus-skin summaries have fewer accepted bins than the corresponding $D_2$ summaries.

\subsection{Nonlinear fourth-harmonic diagnostics}

\begin{figure}[t]
\centering
\includegraphics[width=1.0\textwidth]{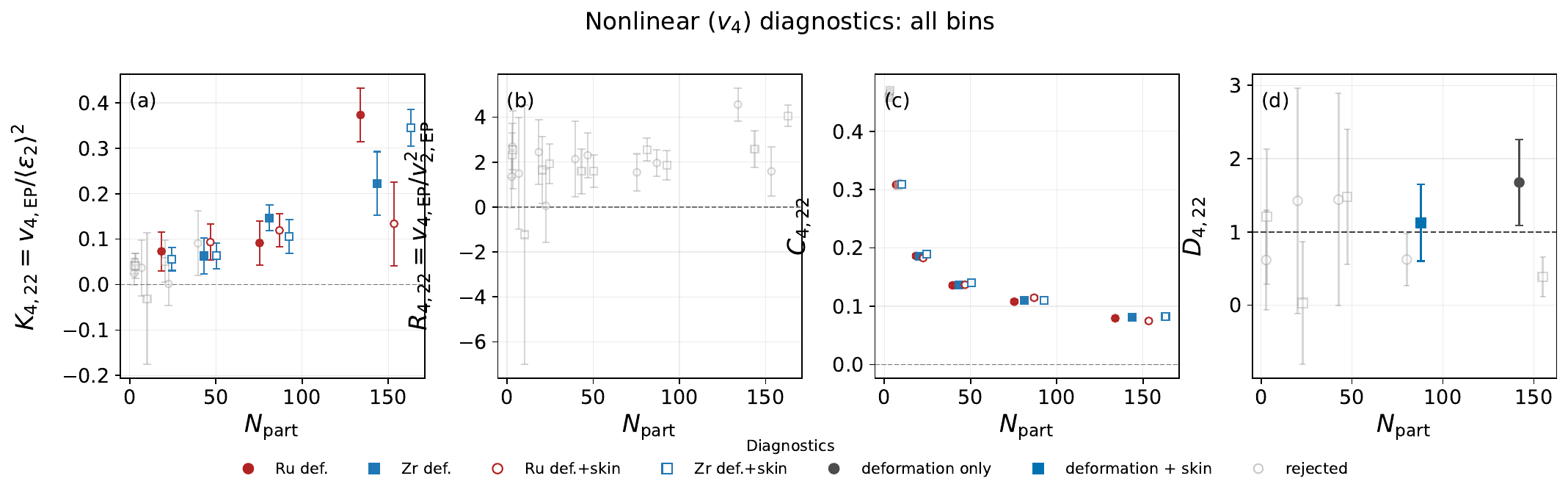}
\caption{Nonlinear fourth-harmonic diagnostics. The present conclusions are based on the directly available eccentricities of n=2 and n=3.}
\label{fig:v4}
\end{figure}

A nonlinear diagnostic is used for the fourth harmonic. The reason is technical and physical: the current eccentricity records contain $\varepsilon_2$ and $\varepsilon_3$ but no explicit $\varepsilon_4$. It is therefore not possible to calculate a direct $\kappa_4=v_4/\varepsilon_4$ or a direct $D_4$ for the analysis. Instead it reports nonlinear diagnostics such as $K_{4,22}=v_4/\varepsilon_2^2$, $R_{4,22}=v_4/v_2^2$ and $C_{4,22}=\langle\cos4(\Psi_4-\Psi_2)\rangle$. These quantities are shown in figure~\ref{fig:v4}. They offer insights into nonlinear response and the present physics conclusions are grounded in the directly available eccentricities for $n=2$ and $n=3$.

This distinction between primary and diagnostic observables is intentional. The $n=2$ and $n=3$ results are directly associated with the eccentricity columns in the input geometry record; the fourth-harmonic results contain both direct and nonlinear response. The same double-ratio method can be extended to $n=4$ in a future study with an explicit $\varepsilon_4$ record. The fourth-harmonic plots are used as targeted diagnostics of the nonlinear stability in the present paper.

\subsection{Null tests}

\begin{figure}[t]
\centering
\includegraphics[width=0.96\textwidth]{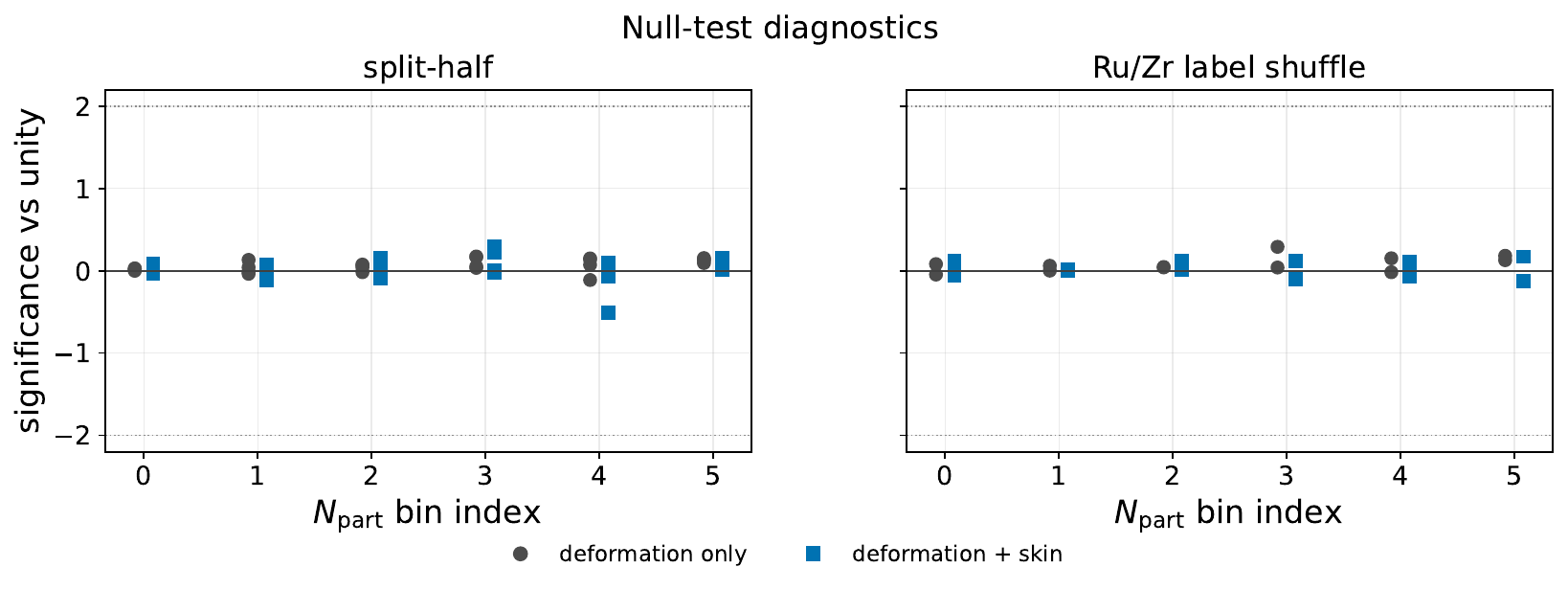}
\caption{Null-test diagnostics from split-half and label-shuffle procedures. The tests confirm the interpretation that the accepted $D_3>1$ trend is not due to a trivial same system split or to random Ru/Zr labelling.}
\label{fig:null}
\end{figure}

A summary of the null tests is given in figure~\ref{fig:null}. Split-half test is used to compare the halves of the same system that are statistically independent, while label-shuffle test breaks the Ru/Zr assignment in similar bins. These tests are not meant to be a replacement for the systematic uncertainty estimate, but they can be used to identify accidental double-ratio structures from statistical partitioning and/or from unstable denominators. The null-test diagnostics are passed for the accepted-bin summaries within the limits of peripheral and low-resolution bins.

The null tests are particularly important for a ratio-of-ratios observable. Both the numerator and the denominator are ratios and a minor denominator instability can create a seemingly clear, but false signal. This risk is minimized if the test is split half consistent or label shuffle consistent. In addition to the systematic variations, these diagnostics provide evidence for understanding that the observed hierarchy for $D_3$ is not a plotting or binning artifact, but a real feature of the output of the AMPT.

\section{Discussion}
\label{sec:discussion}

The primary observation is that there is a harmonic hierarchy. The $D_2$ is consistent with unity for $n=2$ when taking into account the statistical and systematic uncertainties. This means that within AMPT and in the current fixed-$N_{\rm part}$ analysis, the ratio of the elliptic-flow of Ru and Zr is primarily driven by the eccentricity ratio of Ru and Zr. That is, the elliptic response coefficient is roughly the same in the two isobars after the initial difference in eccentricity is divided out.

For $n=3$, the situation is different. In both cases (deformation only and deformation plus skin) the triangular double ratio is systematically greater than one. This means that the final Ru/Zr triangular-flow ratio is larger than expected from the initial Ru/Zr triangularity ratio alone. A possible interpretation is that the correlations between the octupole deformation, participant fluctuations, finite size transport and local density gradient and multiplicity fluctuations affect the effective triangular response. Since triangular flow is more fluctuation driven than elliptic flow, it is not surprising that it is more sensitive to details beyond the mean $\varepsilon_3$ ratio.

The dependence on the neutron skin is a bit more moderate than the harmonic difference between $D_2$ and $D_3$. The positive $D_3-1$ trend is evident in both deformation only and deformation plus skin samples. This does not imply that neutron skin is not relevant. Rather, it means that the particular integrated charged-flow double ratios studied here are not dominated by skin alone. Other observables like multiplicity ratios, mean-$p_T$ fluctuations, identified-particle spectra or charge-sensitive observables could be more sensitive to the proton-neutron density separation.

The two harmonics provide a good contrasting interpretation. The double ratio values $D_2$ lie around the unity value, and $D_3$ values are above unity, all from the same AMPT setup, the same fixed-$N_{\rm part}$ selections, and the same double-ratio construction. This calculation thus defines a particular hierarchy of responses, rather than a common normalization shift. The statistical and systematic treatment gives a quantitative size to the effect: the standard systematic spread places the $D_3$ enhancement at about two standard deviations, while the envelope uncertainty gives an envelope estimate of the variation range across the tested analysis choices. The fixed-$N_{\rm part}$ ensemble-averaged design is the appropriate level for this study because it directly matches the way the Ru/Zr eccentricity and flow ratios are constructed.

The strength of the result lies in its internal consistency. The $D_2$ observables are essentially unity in both flow definitions and both nuclear configurations. The $D_3$ observables remain positive in the event-plane and two-subevent definitions. The pattern withstood the changes in $N_{\rm part}$ binning, peripheral-bin selection, flow $p_T$ range and rapidity/pseudorapidity acceptance. This internal consistency ensures that the result is a meaningful phenomenological prediction for isobar response studies.

\section{Conclusions}
\label{sec:conclusions}

We have studied the transfer of Ru/Zr nuclear-structure differences into final-state anisotropic flow using full-statistics AMPT simulations of $^{96}$Ru+$^{96}$Ru and $^{96}$Zr+$^{96}$Zr collisions at $\sqrt{s_{NN}}=200$ GeV. Four systems were analyzed: deformation-only Ru+Ru, deformation-only Zr+Zr, deformation-plus-skin Ru+Ru and deformation-plus-skin Zr+Zr. The HIJING initialization was modified to include deformed Woods--Saxon nuclei and, when enabled, separate proton and neutron Woods--Saxon parameters. Final state flow analysis was made using the common fixed-$N_{\rm part}$ bins to compare with initial eccentricities.

The main observable introduced in this work was the geometry-normalized double ratio $D_n=(v_n^{\rm Ru}/v_n^{\rm Zr})/(\varepsilon_n^{\rm Ru}/\varepsilon_n^{\rm Zr})$. The elliptic double ratio is consistent with unity, $D_2\simeq1$, indicating that Ru/Zr elliptic-flow differences are largely explained by Ru/Zr eccentricity differences. The triangular double ratio is systematically higher than 1, with stable and quantitatively bounded significance for the different variations of the analysis used. This demonstrates that, in AMPT, the ratio of initial triangular eccentricities is not sufficient to fully describe the triangular response.

The outcome is stable against the considered systematic variations and offers a realistic model prediction for harmonic dependent geometry-to-flow transfer in isobar collisions. Future work should test the same double-ratio observable in independent transport or hydrodynamic frameworks, include direct event-wise $v_n$--$\varepsilon_n$ correlations, and extend the response analysis to identified particles and observables more directly sensitive to neutron skin. Such extensions would help determine how broadly the residual triangular response difference seen here persists across dynamical descriptions of Ru/Zr collisions.

\acknowledgments
The authors would like to thank colleagues and collaborators for discussions on AMPT isobar simulations, nuclear deformation, neutron-skin effects and anisotropic-flow analysis. The numerical calculations were carried out using the modified AMPT/HIJING source code and analyzed using the fixed-$N_{\rm part}$ geometry-to-flow framework described in the text.

\appendix

\section{HIJING--AMPT switch implementation}
\label{app:source}

The deformation and neutron-skin initialization block, which is used to generate the four event samples, is included in the modified HIJING-AMPT source. The deformation switch turns on the deformed Woods–Saxon geometry, the projectile and target switches switch the two nuclei to the same nuclear-structure geometry in a symmetric collision, and the neutron-skin switch chooses either a common density or separate proton and neutron density profiles. The eccentricity-output switch writes the initial geometry record used in the response analysis, and random nuclear orientations are used for the production samples.

The deformation parameters are active for the deformation-only samples while the separation of proton-neutron density is turned off. In the case of the deformation-plus-skin samples the same deformation parameters are employed, but different proton and neutron radii and diffuseness parameters are used. It is important to have the switch logic common to all four samples, since a difference in the residual response is only meaningful if it is not due to a change in the procedure used to generate the events or analyze the data.

\end{document}